%% file: main.tex
\documentclass[11pt,onecolumn,draftcls]{IEEEtran}
\usepackage[isolatin]{inputenc}
\usepackage{listings}
\usepackage{xspace}
\usepackage{here}
\usepackage{graphics, graphicx}
\usepackage{amsmath}
\usepackage{citesort}
\usepackage{fullpage}

\input{macros}

\begin{document}

\title{A Calculus for Sensor Networks}
\author{\authorblockN{Miguel S. Silva\authorrefmark{1},
Francisco Martins\authorrefmark{2}, Lu\'{\i}s Lopes\authorrefmark{1}, and
Jo\~ao Barros\authorrefmark{1}\\[0.2cm]
\authorblockA{\authorrefmark{1}\small Departamento de Ci\^encia de
Computadores \& LIACC\\
Faculdade de Ci\^encias da Universidade do Porto,  Portugal.}\\
\authorblockA{\authorrefmark{2}\small Departamento de Inform\'atica \\
Faculdade de Ci\^encias da Universidade de Lisboa, Portugal.}}
}

\date{}
\maketitle

\input{abstract}
\input{introduction}
\input{calculus}
\input{reductions}
\input{examples}
\input{discussion}
\input{conclusions}
\section*{Acknowledgements}

The authors gratefully acknowledge insightful discussions with Gerhard
Maierbacher (Departamento de Ci\^encia de Computadores, Faculdade de
Ci\^encias, Universidade do Porto).


\end{document}

%% file: macros.tex
\usepackage{listings}

\newcommand{\keyword}[1]{\text{\lstinline|#1|}\,}

\newcommand{\idlek}{\keyword{idle}}
\newcommand{\thisk}{\keyword{this}}
\newcommand{\netk}{\keyword{net}}
\newcommand{\installk}{\keyword{install}}
\newcommand{\sensek}{\keyword{sense}}
\newcommand{\ink}{\keyword{in}}
\newcommand{\offk}{\keyword{off}}
\newcommand{\ifk}{\keyword{if}}
\newcommand{\thenk}{\keyword{then}}
\newcommand{\elsek}{\keyword{else}}
\newcommand{\handlek}{\keyword{handle}}

\newcommand{\inactionn}{\offk}
\newcommand{\inactionp}{\idlek}
\newcommand{\sensor}[5]{[{#1} , {#2}]^{{#3}, {#4}}_{{#5}}}
\newcommand{\sensors}[6]{[{#1} , {#2}, {#3}]^{{#4},{#5}}_{{#6}}}
\newcommand{\sensord}[1][P]{\sensor {#1} M p r b}
\newcommand{\sensorsd}{\sensors H P M p r b}
\newcommand{\tagsensor}[6]{[{#1} , {#2}]^{{#3}, {#4}}_{{#5}}\{{#6}\}}
\newcommand{\tagsensord}[2][P]{\tagsensor {#1} M p r b {#2}}

\newcommand{\moduled}{\{\method{l_i}{\vec x_i}{P_i}\}_{i\in I}}
\newcommand{\method}[3]{{#1} = ({#2})\,{#3}}

\newcommand{\invk}[3]{{#1}.{#2}[{#3}]}
\newcommand{\invkd}{\invk {t}{v}{\vec v}}
\newcommand{\install}[1]{\installk {#1}}
\newcommand{\installd}[1][v]{\install {#1}}
\newcommand{\sense}[2]{\sensek ({#1})\, \ink {#2}}
\newcommand{\sensed}{\sense {\vec x} P}

\newcommand{\parn}{\,\vert\,}
\newcommand{\parp}{\, \vert\, }
\newcommand{\seq}{\;;\;}

\newcommand{\dom}{\operatorname{dom}}

\newcommand{\dist}{\operatorname{d}}
\newcommand{\fn}{\operatorname{fn}}

\newcommand{\energyIn}{\mathsf{c_{in}}}
\newcommand{\energyOut}{\mathsf{c_{out}}}
\newcommand{\methJoin}{+}
\newcommand{\subs}[2]{[{#1}/{#2}]}
\newcommand{\congr}{\equiv}
\newcommand{\reduces}{\rightarrow}

\newcommand{\disprule}[2]{
    #2 \tag{#1}
}

\newcommand{\mkRrule}[1]{\textsc{R-#1}}

\newcommand{\mkSrule}[1]{\textsc{S-#1}}
\newcommand{\Rinstall}{\mkRrule{install}}
\newcommand{\RlocalCall}{\mkRrule{method}}
\newcommand{\RnoLocalCall}{\mkRrule{no-method}}
\newcommand{\RnetCall}{\mkRrule{broadcast}}
\newcommand{\Rrelease}{\mkRrule{release}}
\newcommand{\Rsense}{\mkRrule{sense}}

\newcommand{\Rstructural}{\mkRrule{structural}}
\newcommand{\Rpar}{\mkRrule{parallel}}

\newcommand{\Rnetwork}{\mkRrule{network}}

\newcommand{\Revent}{\mkRrule{event}}

\newcommand{\SmonoidProgram}{\mkSrule{monoid-Program}}
\newcommand{\SmonoidSensor}{\mkSrule{monoid-Sensor}}
\newcommand{\Sbroadcast}{\mkSrule{broadcast}}
\newcommand{\Sbattery}{\mkSrule{bat-exhausted}}
\newcommand{\Sseq}{\mkSrule{idle-seq}}
\newcommand{\Sprogram}{\mkSrule{program-stru}}

\newcommand{\pad}{\;\;}
\newcommand{\Space}[1]{\pad{#1}\pad}
\newcommand{\grmeq}{\Space{::=}}

\newcommand{\grmor}{\;\mid\;}

\newenvironment{myfigure}{
  \begin{figure}[t]\centering\hrulefill\par\vspace{-4ex}}{
    \hrulefill\end{figure}}

\newcommand{\rulespace}{0.3cm}


%% file: abstract.tex
\begin{abstract}

%
We consider the problem of providing a rigorous model for
programming wireless sensor networks. Assuming that collisions,
packet losses, and errors are dealt with at the lower layers
of the protocol stack, we propose a
Calculus for Sensor Networks (CSN) that captures the main
abstractions for programming applications for this class of devices.
%
%
Besides providing the syntax and semantics for the calculus, we show its
expressiveness by providing implementations for several examples of
typical operations on sensor networks.
Also included is a detailed discussion of possible extensions
to CSN that enable the modeling of other important features of these
networks such as sensor state, sampling strategies, and network
security.
\end{abstract}

\textbf{keywords}: Sensor Networks, Ad-Hoc Networks, Ubiquitous Computing, Process-Calculi, Programming Languages.

%% file: introduction.tex

\section{Introduction}
\label{sec:introduction}

\subsection{The Sensor Network Challenge}
Sensor networks, made of tiny, low-cost devices capable of sensing the
physical world and communicating over radio
links~\cite{survey:akyildiz:etal:02}, are significantly different from
other wireless networks: (a) the design of a sensor network is
strongly driven by its particular application, (b) sensor nodes are
highly constrained in terms of power consumption and computational
resources (CPU, memory), and (c) large-scale sensor applications
require self-configuration and distributed software updates without
human intervention.
\begin{figure}[ht]
\centerline{
\includegraphics[width=12cm]{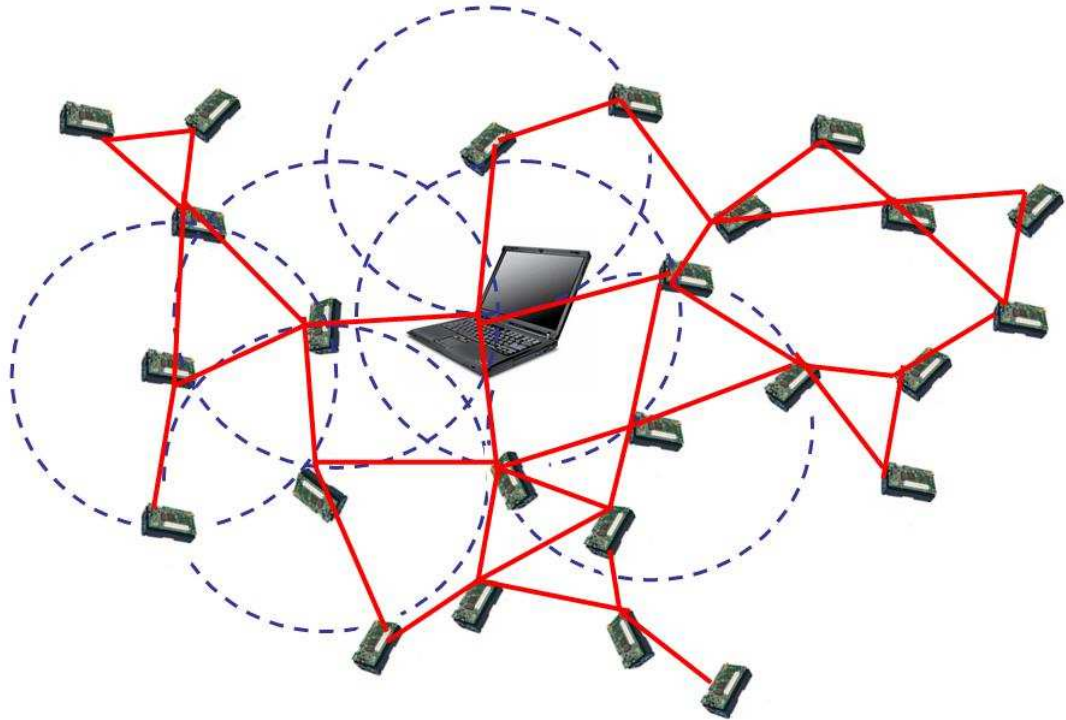}
}
\caption{A wireless sensor network is a collection of small devices
  that, once deployed on a target area, organize themselves in an
  ad-hoc network, collect measurements of a physical process and
  transmit the data over the wireless medium to a data fusion center
  for further processing.}
\label{fig:reachback}
\end{figure}
Previous work on fundamental aspects of wireless sensor networks has
mostly focused on communication-oriented models, in which the sensor
nodes are assumed to store and process the data, coordinate their
transmissions, organize the routing of messages within the network,
and relay the data to a remote receiver (see
\emph{e.g.\@}~\cite{ScaglioneS:02,Hu2004,BarrosS:networkInformationFlow}
and references therein).  Although some of these models provide useful
insights (\emph{e.g.\@} into the connectivity characteristics or the
overall power efficiency of sensor networks) there is a strong need
for formal methods that capture the inherent processing and memory
constraints, and illuminate the massively parallel nature of the
sensor nodes' in-network processing. If well adapted to the specific
characteristics of sensor networks, a formalism of this kind,
specifically a process calculus, is likely to have a strong impact on
the design of operating systems, communication protocols, and
programming languages for this class of distributed systems.

In terms of hardware development, the state-of-the-art is well
represented by a class of multi-purpose sensor nodes called
\emph{motes}\footnote{Trademark of Crossbow Technology,
  Inc.}~\cite{culler:04}, which were originally developed at UC
Berkeley and are being deployed and tested by several research groups
and start-up companies.  In most of the currently available
implementations, the sensor nodes are controlled by module-based
operating systems such as TinyOS~\cite{tinyos} and programming
languages like nesC~\cite{nesc:gay:levis:etal} or
TinyScript/Mat\'{e}~\cite{mate:levis:culler:02}.
In our view, the programming models underlying most of these tools
have one or more of the following drawbacks:
\begin{enumerate}
\item they do not provide a rigorous model (or a calculus) of the
  sensor network at the programming level, which would allow for a
  formal verification of the correctness of programs, among other
  useful analysis;
\item they do not provide a global vision of a sensor network
  application, as a specific distributed application, making it less
  intuitive and error prone for programmers;
\item they require the programs to be installed on each sensor
  individually, something unrealistic for large sensor networks;
\item they do not allow for dynamic re-programming of the network.
\end{enumerate}
Recent middleware developments such as
Deluge~\cite{deluge:hui:culler:04} and
Agilla~\cite{agilla:fok:roman:lu:05} address a few of these drawbacks
by providing higher level programming abstractions on top of TinyOS,
including massive code deployment. Nevertheless, we are still far from
a comprehensive programming solution with strong formal support and
analytical capabilities.

The previous observation motivates us to design a sensor network
programming model from scratch.  Beyond meeting the challenges of
network-wide programming and code deployment, the model should be
capable of producing quantitative information on the amount of
resources required by sensor network programs and protocols, and also
of providing the necessary tools to prove their correctness.

\subsection{Related Work}

Given the distributed and concurrent nature of sensor network
operations, we build our sensor network calculus on thirty years of
experience gathered by concurrency theorists and programming language
designers in pursuit of an adequate formalism and theory for
concurrent systems.
The first steps towards this goal were given by
Milner~\cite{ccs:milner:80} with the development of CCS (Calculus of
Communicating Systems).
CCS describes computations in which concurrent processes may interact
through simple synchronization, without otherwise exchanging
information.
Allowing processes to exchange resources (\emph{e.g.\@}, links,
memory references, sockets, code), besides synchronizing, considerably
increases the expressive power of the formal systems.
Such systems, known as process-calculi, are able to model the mobility
patterns of the resources and thus constitute valuable tools to reason
about concurrent, distributed systems.

The first such system, built on Milner's work, was the
$\pi$-calculus~\cite{pi:milner:parrow:walker:92}.
Later developments of this initial proposal allowed for further
simplification an provided an asynchronous form of the
calculus~\cite{async-pi:honda:tokoro:91,boudol:asynchrony-note:92}.
Since then, several calculi have been proposed to model concurrent
distributed systems and for many there are prototype implementations
of programming languages and run-time systems (\emph{e.g.\@}
Join~\cite{join:fournet:gonthier:96},
TyCO~\cite{dityco:vasco:luis:fernando:98},
X-Klaim~\cite{xklaim:bettini:nicola:pugliese:01}, and Nomadic
Pict~\cite{nomadic:wojciechowski:sewell:00}).

Previous work by Prasad~\cite{broadcast:prasad:91} established the
first process calculus approach to modeling broadcast based systems.
Later work by Ostrovsk\'y, Prasad, and
Taha~\cite{broadcast-high-order:ostrovsky:prasad:taha:02} established
the basis for a higher-order calculus for broadcasting systems.
The focus of this line of work lies in the protocol layer of the
networks, trying to establish an operational semantics and associated
theory that allows assertions to be made about the networks.
More recently, Mezzetti and
Sangiorgi~\cite{wireless:mezzetti:sangiorgi:06} discuss the use of
process calculi to model wireless systems, again focusing on the
details of the lower layers of the protocol stack
(\emph{e.g.\@}~collision avoidance) and establishing an operational
semantics for the networks.

\subsection{Our Contributions}

Our main contribution is a sensor network programming model
based on a process calculus, which we name Calculus of Sensor
Networks (CSN). Our calculus offers the following features that
are specifically tailored for sensor networks:
\begin{itemize}
\item {\it Top-Level Approach}: CSN focuses on programming and
  managing sensor networks and so it assumes that collisions, losses,
  and errors have been dealt with at the lower layers of the protocol
  stack and system architecture (this distinguishes CSN from the
  generic wireless network calculus presented
  in~\cite{wireless:mezzetti:sangiorgi:06});
\item {\it Scalability}: CSN offers the means to provide the sensor
  nodes with self-update and self-configuration abilities, thus
  meeting the challenges of programming and managing a large-scale
  sensor network;
\item {\it Broadcast Communication}: instead of the peer-to-peer
  (unicast) communication of typical process calculi, CSN captures the
  properties of broadcast communication as favored by sensor networks
  (with strong impact on their energy consumption);
\item {\it Ad-hoc Topology}: network topology is not required to be
  programmed in the processes, which would be unrealistic in the case
  of sensor networks;
\item {\it Communication Constraints}: due to the power limitations of
  their wireless interface, the sensor nodes can only communicate with
  their direct neighbors in the network and thus the notion of
  neighborhood of a sensor node, \emph{i.e.}~the set of sensor nodes within
  its communication range, is introduced directly in the calculus;
\item {\it Memory and Processing Constraints}: the typical limitations
  of sensor networks in terms of memory and processing capabilities
  are captured by explicitly modeling the internal processing (or the
  {\it intelligence}) of individual sensors;
\item {\it Local Sensing}: naturally, the sensors are only able to
  pick up local measurements of their environment and thus have
  geographically limited sensitivity.
\end{itemize}

To provide these features, we devise CSN as a two-layer calculus,
offering abstractions for data acquisition, communication, and
processing.
The top layer is formed by a network of sensor nodes
immersed in a scalar or vector field (representing the physical
process captured by the sensor nodes).
The sensor nodes are assumed to be running in parallel.
Each sensor node is composed of a collection of labeled methods, which
we call a module, and that represents the code that can be executed in
the device.
A process is executed in the sensor node as a result of a remote procedure
call on a module by some other sensor or, seen from the point of view
of the callee, as a result of the reception of a message.
Sensor nodes are multithreaded and may share state, for example, in a
tuple-space.
Finally, by adding the notions of position and range, we are able to
capture the nature of broadcast communication and the geographical
limits of the sensor network applications.

The remainder of this paper is structured as follows.
The next section describes the syntax and semantics of the CSN
calculus.
Section~\ref{sec:examples} presents several examples of
functionalities that can be implemented using CSN and that are
commonly required in sensor networks.
In Section~\ref{sec:discussion} we discuss some design options we made
and how we can extend CSN to model other aspects of sensor networks.
%
Finally, Section~\ref{sec:conclusions} presents some conclusions and
directions for future work.


%% file: calculus.tex
\section{The Calculus}
\label{sec:calculus}

This section addresses the syntax and the semantics of the Calculus
for Sensor Networks. 
For simplicity, in the remainder of the paper we will refer to a
sensor node or a sensor device in a network as a \emph{sensor}.
The syntax is provided by the grammar in
Figure~\ref{fig:syntax-sensor}, and the operational semantics is given
by the reduction relation depicted in
Figures~\ref{fig:structural-congruence} and~\ref{fig:reduction-semantics}.

\subsection{Syntax}
\label{sec:syntax}

Let $\vec \alpha$ denote a possible empty sequence $\alpha_1 \dots \alpha_n$ of
elements of the syntactic category $\alpha$.
Assume a countable set of \emph{labels}, ranged over by letter $l$,
used to name methods within modules, and a countable set of
\emph{variables}, disjoint from the set of labels and ranged over by
letter $x$. 
Variables stand for communicated values (\emph{e.g.\@} battery capacity,
position, field measures, modules) in a given program context.

\input{fig-syntax}

The syntax for CNS is found in Figure~\ref{fig:syntax-sensor}.
We explain the syntactic constructs along with their informal,
intuitive semantics.
Refer to the next section for a precise semantics of the calculus.

\emph{Networks} $N$ denote the composition of sensor networks $S$ with
a (scalar or vector) field $F$.
A field is a set of pairs (position, measure) describing the
distribution of some physical quantity (\emph{e.g.\@} temperature,
pressure, humidity) in space.
The position is given in some coordinate system.
Sensors can measure the intensity of the field in their respective
positions.

\emph{Sensor networks} $S$ are flat, unstructured collections of
sensors combined using the parallel composition operator.

A sensor $\sensord$ represents an abstraction of a physical sensing
device and is parametric in its position $p$, describing the location
of the sensor in some coordinate system; its transmission range
specified by the radius $r$ of a circle centered at position $p$; and
its battery capacity $b$.
The position of the sensors may vary with time if the sensor is mobile
in some way.
The transmission range, on the other hand, usually remains constant
over time.
A sensor with the battery exhausted is designated by $\inactionn$.

Inside a sensor there exists a running program~$P$ and a module~$M$.
A module is a collection of methods defined as $l = (\vec x) P$ that the
sensor makes available for internal and for external usage.
A method is identified by label $l$ and defined by an abstraction
$(\vec x) P$: a program $P$ with parameters $\vec x$.
Method names are pairwise distinct within a module.
Mutually recursive method definitions make it possible to represent
infinite behavior.
Intuitively, the collection of methods of a sensor may be interpreted
as the function calls of some tiny operating system installed in the
sensor.

Communication in the sensor network only happens via broadcasting
values from one sensor to its neighborhood: the sensors inside a
circle centered at position $p$ (the position of the sensor) with
radius~$r$.
A broadcast sensor $\tagsensord{S}$ stands for a sensor during
the broadcast phase, having already communicated with sensors $S$. 
While broadcasting, it is fundamental to keep track of the sensors
engaged in communication so far, thus preventing the delivery of the
same message to the same sensor during one broadcasting operation.
Target sensors are collected in the \textit{bag} of the sensor
emitting the message.
Upon finishing the broadcast the bag is emptied out, and the (target)
sensors are released into the network.
This construct is a run-time construct and is available to the
programmer.

Programs are ranged over by $P$.
The $\inactionp$ program denotes a terminated thread.
Method invocation, $\invkd$, selects a method $v$ (with arguments
$\vec v$) either in the local module or broadcasts the request to the
neighborhood sensors, depending whether $t$ is the keyword $\thisk$ or
the keyword $\netk$, respectively.
Program $\sensed$ reads a measure from the surrounding field and binds
it to $\vec x$ within $P$.
Installing or replacing methods in the sensor's module is performed 
using the construct $\installd$.
The calculus also offers a standard form of branching through
the $\ifk\, v\; \thenk\, P\; \elsek P$ construct.

Programs $P$ and $Q$ may be combined in sequence, $P \seq Q$,
or in parallel, $P \parp Q$.
The sequential composition $P \seq Q$ designates a program that first
executes $P$ and then proceeds with the execution of $Q$.
In contrast, $P \parp Q$ represents the simultaneous execution of $P$
and $Q$.
%

\emph{Values} are the data exchanged between sensors and comprise
field measures~$m$, positions~$p$, battery
capacities~$b$, and modules~$M$.
Notice that this is not a higher-order calculus: communicating a
module means the ability to transfer its \emph{code} to, to
retransmit it from, or to install it in a remote sensor.

\subsection{Examples}
\label{sec:examples-1}

Our first example illustrates a network of sensors that sample the
field and broadcast the measured values to a special node known as the
\emph{sink}.
The sink node may be no different from the other sensors in the
network, except that it usually possesses a distinct software module
that allows it to collect and process the values broadcasted in the
network.
The behavior we want to program is the following.
The sink issues a request to the network to sample the field; upon
reception of the request each sensor samples the field at its position
and broadcasts the measured value back to the sink; the sink receives
and processes the values.
An extended version of this example may be found in
Section~\ref{sec:querying}.

The code for the modules of the sensors, \lstinline{MSensor(p, r)},
and for the sink, \lstinline{MSink(p, r)}, is given below.
Both modules are parametric in the position and in the broadcasting
range of each sensor.

As for the module equipping the sensors, it has a method
\lstinline{sample} that, when invoked, propagates the call to its
neighborhood (\lstinline|net.sample[];|), samples the field
(\lstinline{sense x in ...}) and forwards the value to the network
(\lstinline{... net.forward[p,x]}).
Notice that each sensor propagates the original request from the sink.
This is required since in general most of the sensors in the network
will be out of broadcasting range from the sink.
Therefore each sensor echos the request, hopefully covering all
the network.
Message forwarding will be a recurrent pattern found in our examples.
Another method of the sensors' module is \lstinline{forward} that
simply forwards the values from other sensors through the network.

The module for the sink contains a different implementation of the
\lstinline{forward} method, since the sink will gather the values
sent by the sensors and will log them.
Here we leave unspecified the processing done by the
\lstinline{log_position_and_value} program.

The network starts-up with all sensors \lstinline{idle}, except for
the sink that requests a sampling (\lstinline{net.sample[]}).

\begin{lstlisting}
MSensor(p,r) = { sample  = () net.sample[]; sense x in net.forward[p,x]
                 forward = (x,y) net.forward[x,y] }
MSink(p,r)   = { forward = (x,y) log_position_and_value[x,y] }

[net.sample[],MSink(pz,rz)] senS | 
[idle,MSensor(px,rx)] senX | ... | [idle,MSensor(py,ry)] senY
\end{lstlisting}

The next example illustrates the broadcast, the deployment, and 
the installation of code.
The example runs as follows.
The sink node deploys some module in the network
(\lstinline{net.deploy[M]}) and then seals the sensors
(\lstinline{net.seal[]}), henceforth preventing any dynamic
re-programming of the network.
An extended version of the current example may be found in
Section~\ref{sec:sealing-sensors}.

The code for the modules of the sensors and of the sink is given
below.
The module \lstinline{M} is the one we wish to deploy to the network.
It carries the method \lstinline{seal} that forwards the call to the
network and installs a new version of \lstinline{deploy} that does
nothing when executed.

\begin{lstlisting}
MSensor(p,r) = { deploy = (x) net.deploy[x]; install x }
MSink(p,r)   = {}
M            = { seal   = ()  net.seal[]; install { deploy = () idle } }

[net.deploy[M]; net.seal[],MSink(pz,rz)] senS | 
[idle,MSensor(px,rx)] senX | ... | [idle,MSensor(py,ry)] senY
\end{lstlisting}

\subsection{Semantics}
\label{sec:semantics}

The calculus has two name bindings: field sensing and method definitions.
The displayed occurrence of name $x_i$ is a \emph{binding} with
\emph{scope}~$P$ both in  $\sense {x_1, \dots, x_i,
  \dots, x_n} P$ and in $\method l {x_1, \dots,
  x_i, \dots, x_n} P$.
An occurrence of a name is \emph{free} if it is not in the scope of a
binding.
Otherwise, the occurrence of the name is \emph{bound}. 
The set of free names of a sensor $S$ is referred as $\fn(S)$.

\input{fig-stru-cong}

Following Milner~\cite{ccs:milner:80} we present the reduction relation 
with the help of a structural congruence relation.
The structural congruence relation $\congr$, depicted in
Figure~\ref{fig:structural-congruence}, allows for the manipulation of
term structure, adjusting sub-terms to reduce.
The relation is defined as the smallest congruence relation on sensors
(and programs) closed under the rules given in
Figure~\ref{fig:structural-congruence}.

The parallel composition operators for programs and for sensors are
taken to be commutative and associative with $\inactionp$ and
$\inactionn$ as their neutral elements, respectively (\emph{vide} Rules
\SmonoidProgram{} and \SmonoidSensor).
Rule \Sseq{} asserts that $\inactionp$ is also neutral with respect to
sequential composition of programs.
Rule \Sprogram{} incorporates structural congruence for programs
into sensors.
When a sensor is broadcasting a message it uses a bag to collect the
sensors as they become engaged in communication. Rule \Sbroadcast{}
allows for a sensor to start the broadcasting operation.
A terminated sensor is a sensor with insufficient battery capacity for
performing an internal or an external reduction step (\emph{vide} Rule
\Sbattery).

The reduction relation on networks, notation $S, F \reduces S', F$,
describes how sensors $S$ can evolve (reduce) to sensors $S'$, sensing the 
field $F$.
The reduction is defined on top of a reduction relation for sensors,
notation $S \reduces_F S'$, inductively defined by the rules in
Figure~\ref{fig:reduction-semantics}.
The reduction for sensors is parametric on field $F$ and on two
constants $\energyIn$ and $\energyOut$ that represent the amount of
energy consumed when performing internal computation steps
($\energyIn$) and when broadcasting messages ($\energyOut$).

\input{fig-reduction}

Computation inside sensors proceeds by invoking a method (either
local---Rules \RlocalCall{} and \RnoLocalCall---or remote---Rules
\RnetCall{} and \Rrelease), by sensing values (Rule \Rsense), and by
updating the method collection of the sensor (Rule \Rinstall).

The invocation of a local method $l_i$ with arguments $\vec v$ evolves
differently depending on whether or not the definition for $l_i$ is
part of the method collection of the sensor.
Rule \RlocalCall{} describes the invocation of a method from module
$M$, defined as $M(l_i) = (\vec x_i) P_i$.
The result is the program $P_i$ where the values $\vec v$ are bound to
the variables in $\vec x$.
When the definition for $l_i$ is not present in $M$, we have decided 
to actively wait for the definition (see Rule \RnoLocalCall).
Usually invoking an undefined method causes a program to get
\emph{stuck}. 
Typed programming languages use a type system to ensure that there are
no invocations to undefined methods, ruling out all other programs at
compile time.
At run-time, another possible choice would be to simply discard
invocations to undefined methods.
Our choice provides more resilient applications when coupled with the
procedure for deploying code in a sensor network.
We envision that if we invoke a method in the network after some code
has been deployed (see Example~\ref{sec:code-upload}), there may be
some sensors where the method invocation arrives before the deployed
code.
With the semantics we propose, the call actively waits for the code to
be installed.

Sensors communicate with the network by broadcasting messages. 
A message consists of a remote method invocation on unspecified
sensors in the neighborhood of the emitting sensor.
In other words, the messages are not targeted to a particular sensor
(there is no peer-to-peer communication).
The neighborhood of a sensor is defined by its communication radius,
but there is no guarantee that a message broadcasted by a given sensor
arrives at all surrounding sensors.
There might be, for instance, landscape obstacles that prevent two
sensors, otherwise within range, from communicating with each other.
Also, during a broadcast operation the message must only reach
each neighborhood sensor once.
Notice that we are not saying that the same message can not reach the
same sensor multiple times.
In fact it might, but as the result of the echoing of the message in
subsequent broadcast operations.
We model the broadcasting of messages in two stages.
Rule \RnetCall{} invokes method $l_i$ in the remote sensor, provided
that the distance between the emitting and the receiving sensors
is less that the transmission radius ($\dist(p, p') < r$).
The sensor receiving the message is put in the bag of the emitting
sensor, thus preventing multiple deliveries of the same message while
broadcasting.
Observe that the rule does not enforce the interaction with all
sensors in the neighborhood.
Rule \Rrelease{} finishes the broadcast by consuming the operation
($\invk \netk {l_i} {\vec v}$), and by emptying out the contents of
the emitting sensor's bag.
A broadcast operation starts with the application of Rule~\Sbroadcast,
proceeds with multiple (eventually none) applications of
Rule~\RnetCall{} (one for each target sensor), and terminates with the
application of Rule~\Rrelease.

Installing module $M'$ in a sensor with a module $M$, Rule \Rinstall,
amounts to add to $M$ the methods in $M'$ (absent in $M$), and to
replace (in $M$) the methods common to both $M$ and $M'$.
Rigorously, the operation of installing module $M'$ on top of $M$,
denoted $M \methJoin M'$, may be defined as $M \methJoin M' = (M
\setminus M') \cup M'$.
The $\methJoin$ operator is reminiscent of Abadi and Cardelli's
operator for updating methods in their imperative object
calculus~\cite{imperative-object-calculus:abadi:cardelli:95}.

A sensor senses the field in which it is immersed, Rule \Rsense, by
sampling the value of the field $F$ in its position $p$ and, continues
the computation replacing this value for the bound variables $\vec x$
in program~$P$.

Rule \Rpar{} allows reduction to happen in networks of sensors and
Rule \Rstructural{} brings structural congruence into the reduction
relation.



%% file: fig-syntax.tex
\begin{myfigure}
\begin{align*}
  & N \grmeq & & \text{\emph{Network}} & &
  P \grmeq & & \text{\emph{Programs}}\\ 
  & \qquad \;\; S , F & & \text{sensors and field} & &
  \qquad \;\; \inactionp & & \text{idle}\\
  & & & & &
  \quad \grmor P \parp P & & \text{parallel composition}\\
  & S \grmeq & & \text{\emph{Sensors}} & &
  \quad \grmor P \seq P & & \text{sequential composition}\\
  & \qquad \;\; \inactionn & & \text{termination} & &
  \quad \grmor \invkd & & \text{method invocation}\\
  & \quad \grmor N \parn N & & \text{composition} & &
  \quad \grmor \installd & & \text{module update} \\
  & \quad \grmor \sensord & & \text{sensor} & &
  \quad \grmor \sensed & & \text{field sensing}\\
  & \quad \grmor \tagsensord{S} & & \text{broadcast sensor} & &
  \quad \grmor \ifk\ v\ \thenk\ P\ \elsek\ P & & \text{conditional execution}
  \\[0.3cm]
  & M \grmeq & & \text{\emph{Modules}} & &
  v \grmeq & & \text{\emph{Values}}\\ 
  & \qquad \;\; \moduled & & \text{method collection} & &
  \qquad \;\; x & & \text{variable}\\
  & & & & & 
  \quad \grmor m & & \text{field measure}\\
  & t \grmeq & & \text{\emph{Targets}} & &
  \quad \grmor p & & \text{position}\\
  & \qquad \;\; \netk & & \text{broadcast} & &
  \quad \grmor b & & \text{battery capacity}\\
  & \quad \grmor \thisk & & \text{local} & &
  \quad \grmor M & & \text{module}
\end{align*}
\caption{The syntax of CSN.}
\label{fig:syntax-sensor}
\end{myfigure}


%% file: fig-stru-cong.tex
\begin{myfigure}
  \begin{gather*}
    \tag{\SmonoidProgram}
    P_1 \parp P_2 \congr P_2 \parp P_1,
    \qquad
    P \parp \inactionp \congr P,
    \qquad
    P_1 \parp (P_2 \parp P_3) \congr (P_1 \parp P_2) \parp P_3
    \\
    \tag{\SmonoidSensor}
    S_1 \parn S_2 \congr S_2 \parn S_1,
    \qquad
    S \parn \inactionn \congr S,
    \qquad
    S_1 \parn (S_2 \parn S_3)  \congr (S_1 \parn S_2) \parn S_3
    \\
    \tag{\Sseq, \Sprogram}
    \inactionp \seq P \congr P
    \qquad \qquad
    \frac{
      P_1 \congr P_2
    }{
      \sensord[P_1] \congr \sensord[P_2]
    }
    \\
    \tag{\Sbroadcast, \Sbattery}
    \sensord \congr \tagsensord[P]{\inactionn}
    \qquad \qquad
    \frac{
      b < \max(\energyIn, \energyOut)
    }{
      \sensord[P] \congr \inactionn
    }
  \end{gather*}
\caption{Structural congruence for processes and sensors.}
\label{fig:structural-congruence}
\end{myfigure}


%% file: fig-reduction.tex
\input{fig-oper-semantic}


%% file: fig-oper-semantic.tex
\begin{myfigure}
  \begin{gather*}
    \disprule{\RlocalCall}
    {
      \frac{
        M (l_i) = (\vec x_i) P_i
        \qquad
        b \geq \energyIn
      }
      {
        \sensord[\invk \thisk {l_i} {\vec v} \seq P_1 \parp P_2]
        \reduces_F 
        \sensor {P_i \subs {\vec v} {\vec x_i} \seq P_1 \parp P_2} M
        p r {b - \energyIn} 
      }
    }
    \\[\rulespace]
    \disprule{\RnoLocalCall}
    {
      \frac{
        l_i \not \in \dom(M)
      }
      {
        \sensord[\invk \thisk {l_i} {\vec v} \seq P_1 \parp P_2]
        \reduces_F
        \sensor {\invk \thisk {l_i} {\vec v} \seq P_1 \parp P_2} M
        p r b
      }
    }
    \\[\rulespace] 
    \disprule{\RnetCall}
    {
      \frac{
        \dist (p, p') < r 
        \qquad
        b \geq \energyOut
      }
      {
        \begin{array}{l}
        \tagsensord[\invk \netk {l_i} {\vec v} \seq P_1 \parp P_2]{S} \parn
          \sensor{P'}{M'}{p'}{r'}{b'}  
          \reduces_F\\
          \hspace{3cm}
          \tagsensord[\invk \netk {l_i} {\vec v} \seq P_1 \parp P_2]
          {S \parn \sensor{P' \parp \invk {\thisk} {l_i} {\vec
                v}}{M'}{p'}{r'}{b'}}  
        \end{array}
      }
    }
    \\[\rulespace]
    \disprule{\Rrelease}
    {
        \tagsensord[\invk \netk {l_i} {\vec v} \seq P_1 \parp P_2]{S}
        \reduces_F
        \sensor {P_1 \parp P_2} M p r {b - \energyOut} \parn S
    }
    \\[\rulespace]
    \disprule{\Rinstall}
    {
      \frac{
        b \geq \energyIn
      }
      {
        \sensord[{\installd[M']} \seq P_1 \parp P_2] 
        \reduces_F
        \sensor {P_1 \parp P_2}
                {M \methJoin M'} p r {b - \energyIn}
      }
    }
    \\[\rulespace]
    \disprule{\Rsense}
    {
      \frac{
        b \geq \energyIn
      }
      {
        \sensord[\sensed \seq P_1 \parp P_2] 
        \reduces_F
        \sensor {P \subs {F(p)} {\vec x} \seq P_1 \parp P_2} M
        p r {b - \energyIn}
      }
    }
    \\[\rulespace]
    \tag{\Rpar, \Rstructural}
    \frac{
      S_1 \reduces_F S_2
    }
    {
      S \parn S_1 
      \reduces_F
      S \parn S_2
    }
    \qquad
    \frac{
      S_1 \congr S_2 \quad\;\; 
      S_2 \reduces_F S_3 \quad\;\;
      S_3 \congr S_4
    }
    {
      S_1
      \reduces_F 
      S_4
    }
    \\[\rulespace]
    \disprule{\Rnetwork}
    {
      \frac{
        S \reduces_F S'
      }
      {
        S, F 
        \reduces
        S', F
      }
    }
  \end{gather*}
\caption{Reduction semantics for processes and networks.}
\label{fig:reduction-semantics}
\end{myfigure}


%% file: reductions.tex
\subsection{The Operational Semantics Illustrated}
\label{sec:illustrated}

To illustrate the operational semantics of CNS, 
we present the reduction steps for the examples discussed 
at the end of Section~\ref{sec:examples-1}.
During reduction we suppress the side annotations when writing the
sensors.
Due to space constraints we consider a rather simple network with just
the sink and another sensor.
\begin{lstlisting}
[net.sample[],MSink(pz,rz)] | [idle,MSensor(px,rx)] 
\end{lstlisting}
We assume that the sensor is within range from the sink and vice-versa.
This network may reduce as follows:
 
\begin{align*}
& \text{\lstinline{[net.sample[],MSink(pz,rz)] | [idle,MSensor(px,rx)]}} \equiv 
\tag{\Sbroadcast} \\
& \text{\lstinline{[net.sample[],MSink(pz,rz)]\{off\} | [idle,MSensor(px,rx)]}} \rightarrow \equiv
\tag{$\dist(p,p_1)<r$, \RnetCall, \SmonoidSensor} \\
& \text{\lstinline{[net.sample[],MSink(pz,rz)] \{[this.sample[] || idle,MSensor(px,rx)]\}}} \rightarrow \equiv  
\tag{\Rrelease, \SmonoidProgram} \\
& \text{\lstinline{[idle,MSink(pz,rz)] | [this.sample[],MSensor(px,rx)]}}  \rightarrow  
\tag{\RlocalCall} \\
& \text{\lstinline{[idle,MSink(pz,rz)] | [net.sample[]; sense x in net.forward[px,x],MSensor(px,rx)]}}  \equiv 
\tag{\Sbroadcast} \\
& \text{\lstinline{[idle,MSink(pz,rz)] | [net.sample[]; sense x in net.forward[px,x],MSensor(px,rx)]\{off\}}}  \rightarrow   \tag{\Rrelease} \\
& \text{\lstinline{[idle,MSink(pz,rz)] | [sense x in net.forward[px,x],MSensor(px,rx)]}}  \rightarrow   
\tag{\Rsense} \\
& \text{\lstinline{[idle,MSink(pz,rz)] | [net.forward[px,fz(px)], MSensor(px,rx)]}}  \equiv    
\tag{\SmonoidSensor} \\
& \text{\lstinline{[net.forward[px,fz(px)], MSensor(px,rx)] | [idle,MSink(pz,rz)]}} \equiv    
\tag{\Sbroadcast} \\
& \text{\lstinline{[net.forward[px,fz(px)], MSensor(px,rx)]\{off\} | [idle,MSink(pz,rz)]}} \rightarrow \equiv   
\tag{$\dist(p_1,p)<r_1$, \RnetCall, \SmonoidSensor} 
\end{align*}
\begin{align*}
& \text{\lstinline{[net.forward[px,fz(px)], MSensor(px,rx)]\{[this.forward[px,fz(px)] || idle, MSink(pz,rz)]\}}} \rightarrow \equiv \tag{\Rrelease, \SmonoidProgram} \\
& \text{\lstinline{[idle, MSensor(px,rx)] | [this.forward[px,fz(px)], MSink(pz,rz)]}} \rightarrow 
\tag{\RlocalCall} \\
& \text{\lstinline{[idle, MSensor(px,rx)] | [log_position_and_value[px,fz(px)], MSink(pz,rz)]}}  \equiv 
\tag{\SmonoidSensor} \\
& \text{\lstinline{[log_position_and_value[px,fz(px)] | [idle, MSensor(px,rx)], MSink(pz,rz)]}}  
\end{align*}
So, after these reduction steps the sink gets the field values from
the sensor at position \lstinline{px} and logs them.
The sensor at \lstinline{px} is idle waiting for further interaction. 

Following we present the reduction step for our second (and last) example
of Section~\ref{sec:examples-1} where we illustrate the broadcast,
the deployment, and the installation of code.
Again, due to space restrictions, we use a very simple network with
just the sink and another sensor, both within reach of each other.
\begin{lstlisting}
[net.deploy[M]; net.seal[],MSink(pz,rz)] | [idle,MSensor(px,rx)]
\end{lstlisting}
This network may reduce as follows:
\begin{align*}
& \text{\lstinline{[net.deploy[M]; net.seal[],MSink(pz,rz)] | [idle,MSensor(px,rx)]}} \equiv 
\tag {\Sbroadcast}  
\\
& \text{\lstinline{[net.deploy[M]; net.seal[],MSink(pz,rz)]\{off\} | [idle,MSensor(px,rx)]}} \rightarrow \equiv  
\tag {$\dist(p,p_1)<r$, \RnetCall, \SmonoidSensor}  
\\
& \text{\lstinline{[net.deploy[M]; net.seal[],MSink(pz,rz)] | \{[this.deploy[M] || idle, MSensor(px,rx)]\}}} \rightarrow 
\equiv \tag{\Rrelease, \SmonoidProgram} 
\\
& \text{\lstinline{[net.seal[],MSink(pz,rz)] | [this.deploy[M], MSensor(px,rx)]}} \rightarrow 
\tag{\RlocalCall}
\\
& \text{\lstinline{[net.seal[],MSink(pz,rz)] | [net.deploy[M]; install M, MSensor(px,rx)]}} \equiv 
\tag{\Sbroadcast}
\\
& \text{\lstinline{[net.seal[],MSink(pz,rz)] | [net.deploy[M]; install M, MSensor(px,rx)]\{off\}}} \rightarrow \equiv 
\tag{\Rrelease, \SmonoidSensor}
\\
& \text{\lstinline{[net.seal[],MSink(pz,rz)] | [install M, MSensor(px,rx)]}} \rightarrow 
\tag{\Rinstall} 
\\
& \text{\lstinline{[net.seal[],MSink(pz,rz)] | [idle, MSensor(px,rx)+M]}} \equiv
\tag{\Sbroadcast}
\\ 
& \text{\lstinline{[net.seal[],MSink(pz,rz)]\{off\} | [idle, MSensor(px,rx)+M]}} \rightarrow \equiv
\tag{$\dist(p,p_1)<r$, \RnetCall, \SmonoidSensor} \\
& \text{\lstinline{[net.seal[],MSink(pz,rz)] \{[this.seal[] || idle, MSensor(px,rx)+M]\}}} \rightarrow \equiv 
\tag{\Rrelease, \SmonoidProgram} 
\end{align*}
\begin{align*}
& \text{\lstinline{[idle,MSink(pz,rz)] | [this.seal[], MSensor(px,rx)+M]}} \rightarrow 
\tag{\RlocalCall} 
\\
& \text{\lstinline{[idle,MSink(pz,rz)] | [net.seal[]; install \{deploy = () idle\}, MSensor(px,rx)+M]}} \equiv  
\tag{\Sbroadcast}
\\
& \text{\lstinline{[idle,MSink(pz,rz)] | [net.seal[]; install \{deploy = () idle\}, MSensor(px,rx)+M]\{off\}}} \rightarrow \equiv 
\tag{\Rrelease, \SmonoidSensor}
\\
& \text{\lstinline{[idle,MSink(pz,rz)] | [install \{deploy = () idle\}, MSensor(px,rx)+M]}} \rightarrow 
\tag{\Rinstall} 
\\
& \text{\lstinline{[idle,MSink(pz,rz)] | [idle, MSensor(px,rx)+M+\{deploy = () idle\}]}} 
\end{align*}

After these reductions, the sink is idle after deploying the code to
the sensor at \lstinline{pz}.
The sensor at \lstinline{pz} is also idle, waiting for interaction, but
with the code for the module \lstinline{M} installed and with the
\lstinline{deploy} method disabled.


%% file: examples.tex
\section{Programming Examples}
\label{sec:examples}

In this section we present some examples, programmed in CSN, of
typical operations performed on networks of sensors.
Our goal is to show the expressiveness of the CSN calculus just
presented and also to identify some other aspects of these networks
that may be interesting to model.
In the following examples, we denote as \lstinline{MSensor} and
\lstinline{MSink} the modules installed in any of the anonymous sensors
in the network and the modules installed in the sink, respectively.
Note also that all sensors are assumed to have a builtin method,
\lstinline{deploy}, that is responsible for installing new modules.
The intuition is that this method is part of the tiny operating
system that allows sensors to react when first placed in the field.
Finally, we assume in these small examples that the network layer 
supports \emph{scoped flooding}.
We shall see in the next section that this can be supported via
software with the inclusion of state in sensors.
 
\subsection{Ping}
\label{sec:ping}

We start with a very simple \lstinline{ping} program.
Each sensor has a \lstinline{ping} method that when invoked calls a
method \lstinline{forward} in the network with its position and
battery charge as arguments.
When the method \lstinline{forward} is invoked by a sensor in the
network, it just triggers another call to \lstinline{forward} in the
network.
The sink has a distinct implementation of this method.
Any incomming invocation logs the position and battery values given as
arguments.
So, the overall result of the call \lstinline{net.ping[]} in the sink
is that all reachable sensors in the network will, in principle,
receive this call and will flood the network with their positions and
battery charge values.
These values eventually reach the sink and get logged.
 
\begin{lstlisting}
  MSensor(p,b) = {
    ping           = ()    net.forward[p,b]; net.ping[]
    forward        = (x,y) net.forward[x,y]
  }
  MSink(p,b)   = {
    forward        = (x,y) log_position_and_power[x,y]
  }
  [net.ping[], MSink(pz,bz)] senS | 
  [idle, MSensor(px,bx)] senX  | ... | [idle, MSensor(py,by)] senY
\end{lstlisting}

\subsection{Querying}
\label{sec:querying}

This example shows how we can program a network with a sink that
periodically queries the network for the readings of the sensors.
Each sensor has a \lstinline{sample} method that samples the field using
the \sensek{} construct and calls the method \lstinline{forward} in the
neighbourhood with its position and the value sampled as arguments.
The call then queries the neighbourhood recursively with a replica of
the original call.
The original call is, of course, made from the sink, which has a method
\lstinline{start_sample} that calls the method \lstinline{sample} in the
network within a cycle.
Note that, if the sink had a method named \lstinline{sample} instead of
\lstinline{start_sample}, it might get a call to \lstinline{sample} from
elsewhere in the network that could interfere with the sampling
control cycle.

\begin{lstlisting}
  MSensor(p)   = {
    sample         = ()     sense (x) in net.forward[p, x]; net.sample[]
    forward        = (x, y) net.forward[x,y]
  }
  MSink(p)     = {
    start_sample   = ()     net.sample[]; this.start_sample[]
    forward        = (x, y) log_position_and_value[x,y]
  }
  [this.start_sample[], MSink(pz)] senS | 
  [idle, MSensor(px)] senX  | ... | [idle, MSensor(py)] senY
\end{lstlisting}

\subsection{Polling}
\label{sec:polling}

In this example the cycle of the sampling is done in each sensor,
instead of in the sink, as in the previous example.
The sink just invokes the method \lstinline{start_sample} once.
This method propagates the call through the network and invokes
\lstinline{sample}, for each sensor.
This method samples the field, within a cycle, and forwards the
result to the network.
This implementation requires less broadcasts than the previous one as
the sink only has to call \lstinline{start_sample} on the network
once.
On the other hand, it increases the amount of processing per sensor.

\begin{lstlisting}
  MSensor(p)   = {
    start_sample   = ()     net.start_sample[]; this.sample[]
    sample         = ()     sense (x) in net.forward[p, x]; this.sample[]
    forward        = (x, y) net.forward[x, y]
  }
  MSink(p)     = {
    forward        = (x, y) log_position_and_value[x,y]
  }
  [net.start_example[], MSink(pz)] senS | 
  [idle, MSensor(px)] senX  | ... | [idle, MSensor(py)] senY
\end{lstlisting}

\subsection{Code deployment}
\label{sec:code-upload}

The above examples assume we have some means of deploying the code to
the sensors.
In this example we address this problem and show how it can be
programmed in CSN.
The code we wish to deploy and execute is the same as the one in the
previous example.
To achieve this goal, the sink first calls the \lstinline{deploy} method
on the network to install the new module with the methods
\lstinline{start_sample}, \lstinline{sample} and \lstinline{forward} as above.
This call recursively deploys the code to the sensors in the network.
The sink then calls \lstinline{start_sample} to start the sampling,
again as above, and waits for the forwarded results on the method
\lstinline{forward}.

\begin{lstlisting}
  MSensor(p)   = {
    deploy         = (x)    install x; net.deploy[x]
  }
  MSink(p)     = {
    forward        = (x,y)  log_position_and_value[x,y]
  }
  [net.deploy[{
    start_sample   = ()     net.start_sample[]; this.sample[]
    sample         = ()     sense (x) in net.forward[p, x]; this.sample[]
    forward        = (x, y) net.forward[x, y]
   }];
   net.start_sample[], MSink(pz)] senS | 
  [idle, MSensor(px)] senX  | ... | [idle, MSensor(py)] senY
\end{lstlisting}

A refined version of this code, one that avoids the
\lstinline{start_sample} method completely, can be programmed.
Here, we deploy the code for all sensors by sending methods
\lstinline{sample} and \lstinline{forward} to all the sensors in the
network by invoking \lstinline{deploy}.
Once deployed, the code is activated with a call to \lstinline{sample} in
the sink, instead of using the \lstinline{start_sample} method as
above.
\begin{lstlisting}
  MSensor(p)   = {
    deploy         = (x)   install x; net.deploy[x]
  }
  MSink(p)     = {
    forward        = (x,y) log_position_and_value[x,y]
  }
  [net.deploy[{
    sample         = ()    net.sample[];
                           install {sample = () sense (x) in net.forward[p, x]; 
                                                this.sample[]};
                           this.sample[]
    forward        = (x, y) net.forward[x, y]
   }];
   net.sample[], MSink(pz)] senS | 
  [idle, MSensor(px)] senX  | ... | [idle, MSensor(py)] senY
\end{lstlisting}
Notice that the implementation of the method \lstinline{sample} has
changed.
Here, when the method is executed for the first time at each sensor,
it starts by propagating the call to its neighborhood and then, it
changes itself through an \installk{} call.
The newly installed code of \lstinline{sample} is the same as the one
in the first implementation of the example.
The method then continues to execute and calls the new version of
\lstinline{sample}, which starts sampling the field and forwarding
values.
%

\subsection{Sealing sensors}
\label{sec:sealing-sensors}

This example shows how we can install a sensor network with a module
that contains a method, \lstinline{seal}, that prevents any further
dynamic re-programming of the sensors, preventing anyone from
tampering with the installed code.
The module also contains a method, \lstinline{unseal} that restores the 
original \lstinline{deploy} method, thus allowing dynamic re-programming again.
The sink just installs the module containning these methods in the
network by broadcasting a method call to \lstinline{deploy}.
Each sensor that receives the call, installs the module and floods the
neighborhood with a replica of the call.
Another message by the sink then replaces the \lstinline{deploy} method
itself and re-implements it to \idlek{}. 
This prevents any further instalation of software in the sensors and
thus effectively seals the network from external interaction other
than the one allowed by the remainder of the methods in the modules of
the sensors.
 
\begin{lstlisting}
  MSensor      = {
    deploy     = (x) install x; net.deploy[x]
  }
  MSink        = { }
  [net.deploy[{ 
    seal       = ()  install {deploy = ()  idle}
    unseal     = ()  install {deploy = (x) install x; net.deploy[x]}
   }];
   net.seal[], MSink] senS | 
  [idle, MSensor] senX  | ... | [idle, MSensor] senY
\end{lstlisting}


%% file: discussion.tex
\section{Discussion}
\label{sec:discussion}

In the previous sections, we focused our attention on the programming
issues of a sensor network and presented a core calculus that is
expressive enough to model fundamental operations such as local
broadcast of messages, local sensing of the environment, and software
module updates.
CSN allows the global modeling of sensor networks in the sense that it
allows us to design and implement sensor network applications as
large-scale distributed applications, rather than giving the programmer
a sensor-by-sensor view of the programming task.
It also provides the tools to manage running sensor networks,
namely through the use of the software deployment capabilities.

There are other important features of sensor networks that we
consciously left out of CSN.
In the sequel we discuss some of these features and sketch some ideas
of how we would include support for them.
 
\paragraph{State}

From a programming point of view, adding state to sensors is essential.
Sensors have some limited computational capabilities and may perform
some data processing before sending it to the sink.
This processing assumes that the sensor is capable of buffering data
and thus maintain some state.
In a way, CSN sensors have state. Indeed, the atributes $p$, $b$, and
$r$ may be viewed as sensor state.
Since these are characteristic of each sensor and are usually
controlled at the hardware level, we chose to represent this state as
parameters of the sensors.
The programmer may read these values at any time through builtin
method calls but any change to this data is performed transparently
for the programmer by the hardware or operating system.
As we mentioned before, it is clear that the value of $b$ changes
with time.
The position $p$ may also change with time if we envision our sensors
endowed with some form of mobility (\emph{e.g.\@}, sensors dropped in the
atmosphere or flowing in the ocean).

To allow for a more systematic extension of our sensors with state
variables we can assume that each sensor has a heap $H$ where the values
of these variables are stored: $\sensorsd$.
The model chosen for this heap is orthogonal to our sensor calculus
and for this discussion we assume that we enrich the values $v$ of the
language with a set of \emph{keys}, ranged over by $k$.
Our heap may thus be defined as a map $H$ from $keys$ into $values$.
Intuitively, we can think of it as an associative memory with the
usual built-in operations \lstinline{put}, \lstinline{get},
\lstinline{lookup}, and \lstinline{hash}.
Programs running in the sensors may share state by exchanging keys.
We assume also that these operations are atomic and thus no race
conditions can arise.

With this basic model for a heap we can re-implement the \emph{Ping}
example from Section~\ref{sec:ping} with \emph{scoped flooding} thus
eliminating echos by software.
We do this by associating a unique key to each remote procedure call
broadcast to the network.
This key is created through the built-in \lstinline{hash} function that
takes as arguments the position \lstinline{p} and the battery
\lstinline{b} of the sensor.
%
%
Each sensor, after receiving a call to \lstinline{ping}, propagates
the call to its neighborhood and generates a new key to send, with its
position and battery charge, in a \lstinline{forward} call.
Then, it stores the key in its heap to avoid forwarding its own
\lstinline{forward} call.
On the other hand, each time a sensor receives a call to 
\lstinline{forward}, it checks whether it has the key associated to the call 
in its heap.
If so, it does nothing.
If not, it forwards the call and stores the key in the heap, to avoid
future re-transmission.

\begin{lstlisting}
  MSensor(p, b) = {
    ping       = ()        net.ping[]; 
                           let k = hash[p,b] in net.forward[p, b, k];
                           put[k,_]
    forward    = (x, y, k) if (!lookup[k]) then 
                              net.forward[x, y, k]; 
                              put[k,_]
  }
  MSink(p, b)   = {...}
  [net.ping[], MSink] senS | 
  [idle, MSensor(px,bx)] senX  | ... | [idle, MSensor(py,by)] senY
\end{lstlisting}


\paragraph{Events}
Another characteristic of sensors is their \emph{modus operandi}. 
Some sensors sample the field as a result of instructions
implemented in the software that controls them. 
Such is the case with CSN sensors.
The programmer is responsible for controlling the sensing activity of
the sensor network.
It is of course possible for sensor nodes to be activated in different
ways. 
For example, some may have their sensing routines implemented at
hardware or operating system level and thus not directly controllable
by the programmer.
Such classes of sensor nodes tipically sample the field periodically
and are activated when a given condition arises (\emph{e.g.\@}, a temperature
above or below a given threshold, the detection of CO$_2$ above a given
threshold, the detection of a strong source of infrared light).
The way in which certain environmental conditions or \emph{events} can
activate the sensor is by triggering the execution of a handler
procedure that processes the event.
Support for this kind of event-driven sensors in CSN could be
achieved by assuming that each sensor has a builtin handler procedure,
say \lstinline{handle}, for such events.
The handler procedure, when activated, receives the value of the field that 
triggered the event.
Note that, from the point of view of the sensor, the occurrence of
such an event is equivalent to the deployment of a method invocation
\lstinline{this.handle[v]} in its processing core, where \lstinline{v} is
the field value associated with the event.
The sensor has no control over this deployment, but may be programmed
to react in different ways to these calls, by providing adequate
implementations of the \lstinline{handle} routine.
The events could be included in the semantics given in
Section~\ref{sec:semantics} with the following rule:
\begin{gather*}
    \disprule{\Revent}
    {
        \sensord[P] 
        \reduces_F
        \sensor {\invk \thisk {\handlek} {F(p)} \parp P} M
        p r b
    }
\end{gather*}
As in the case of the builtin method for code deployment, the handler
could be programmed to change the behavior of the network in the
presence of events.
One could envision the default handler as \lstinline{handle = (x) idle}, 
which ignores all events.
Then, we could change this default behavior so that an event triggers
an alarm that gets sent to the sink.
A possible implementation of such a dynamic re-programming of the network
default handlers can be seen in the code below.

\begin{lstlisting}
  MSensor(p)  = { handle  = (x) idle }
  MSink(p)    = { handle  = (x) idle }
  [net.deploy[{ 
       handle = (x)   net.alarm[p,x]
       alarm  = (x,y) net.alarm[x,y] 
   }];
   install {alarm  = (x,y) sing_bell[x,y]},
   MSink(pz)] senS | 
  [idle, MSensor(px)] senX  | ... | [idle, MSensor(py)] senY
\end{lstlisting}

where the default implementation of the \lstinline{handle} procedure is
superseded by one that eventualy triggers an alarm in the sink.
 
More complex behavior could be modeled for sensors that take multiple
readings, with a handler associated with each event.

\paragraph{Security}
Finally, another issue that is of outmost importance in the management
of sensor networks is security.
It is important to note that many potential applications of sensor 
networks are in high risk situations.
Examples may be the monitorization of ecological disaster areas, 
volcanic or sismic activity, and radiation levels in contaminated areas.
Secure access to data is fundamental to establish its credibility and 
for correctly assessing risks in the management of such episodes.  
In CSN we have not taken security issues into consideration. This was
not our goal at this time.
However, one feature of the calculus may provide interesting solutions
for the future.
In fact, in CSN, all computation within a sensor results from an
invocation of methods in the modules of a sensor, either originating
in the network or from within the sensor.
In a sense the modules $M$ of the sensor work as a firewall that can be used
to control incomming messages and implement security protocols.
Thus, all remote method invocations and software updates might first
be validated locally with methods of the sensor's modules and
only then the actions would be performed.
The idea of equipping sensors, or in general \emph{domains}, with some
kind of membrane that filters all the interactions with the surrounding
network has been explored, for instance,
in~\cite{boudol:generic-membrane-model}, in the
M-calculus~\cite{schmitt.stefani:higer-order-process-calculus}, in the
Kell calculus~\cite{stefani:calculus-kells}, in the Brane
calculi~\cite{cardelli:brane-calculi}, in Miko~\cite{martins.etal:miko},
and in~\cite{gorla.hennessy.sassone:security-policies-membranes-gc}.
One possible development is to incorporate some features of the
\emph{membrane model} into CSN.
The current formulation of the calculus also assumes that all methods
in the module $M$ of a sensor $\sensord$ are visible from the network.
It is possible to implement an access policy to methods in such a way
that some methods are private to the sensor, \emph{i.e.}, can only be
invoked from within the sensor.
This allows, for example, the complete encapsulation of the state of
the sensor.


%% file: conclusions.tex
\section{Conclusions and Future Work}
\label{sec:conclusions}

Aiming at providing large-scale sensor networks with a rigorous and adequate 
programming model (upon which operating systems and high-level programming
languages can be built), we presented CSN --- a Calculus for 
Sensor Networks, developed specifically for this class of distributed systems. 

After identifying the necessary sensing, processing, and wireless
broadcasting features of the calculus, we opted to base our work on a
top-layer abstraction of physical and link layer communication issues
(in contrast with previous work on wireless network
calculi~\cite{wireless:mezzetti:sangiorgi:06,broadcast:prasad:91}),
thus focusing on the system requirements for programming network-wide
applications.  This approach resulted in the CSN syntax and semantics,
whose expressiveness we illustrated through a series of
implementations of typical operations in sensor networks. Also
included was a detailed discussion of possible extensions to CSN to
account for other important properties of sensors such as state,
sampling strategies, and security.

As part of our ongoing efforts, we are currently using CSN to establish
a mathematical framework for reasoning 
about sensor networks.
One major objective of this work consists in providing formal proofs
of correctness for data gathering protocols that are commonly used in 
current sensor networks and whose performance and reliability 
has so far only been evaluated through computer simulations and
ad-hoc experiments.

From a  more practical point of view, the focus will be set on the development
of a prototype implementation of CSN. This prototype 
will be used to emulate the behavior of sensor networks by software 
and, ultimately, to port the programming model to a natural 
development architecture for sensor network applications.
